# Structural and electronic properties of ultrathin FeSe films grown on Bi$_2$Se$_3$(0001) studied by STM/STS

U. R. Singh,[*] J. Warmuth, V. Markmann, J. Wiebe,[§] and R. Wiesendanger

*Department of Physics, University of Hamburg, D-20355 Hamburg, Germany*

## Abstract

We report scanning tunnelling microscopy and spectroscopy (STM/STS) studies on one and two unit cell (UC) high FeSe thin films grown on Bi$_2$Se$_3$(0001). In our thin films, we find the tetragonal phase of FeSe and dumb-bell shaped defects oriented along Se-Se bond directions. In addition, we observe striped moiré patterns with a periodicity of (7.3±0.1) nm generated by the mismatch between the FeSe lattice and the Bi$_2$Se$_3$ lattice. We could not find any signature of a superconducting gap in the tunneling spectra measured on the surface of one and two UC thick islands of FeSe down to 6.5 K. The spectra rather show an asymmetric behavior across and a finite density of states at the Fermi level ($E_F$) resembling those taken in the normal state of bulk FeSe.

## Introduction

Several years after the discovery of iron-based superconductors, the iron-chalcogenide superconductors have been found to exhibit the simplest crystal and chemical structures while the superconducting transition temperature ($T_C$) can be as high as ~8.5 K for bulk FeSe[1,2] and ~14 K for FeSe$_x$Te$_{1-x}$.[2,3] In the former case, $T_C$ can be enhanced up to 35 K by applying pressure[4] and up to 45 K by doping different charge carriers[5]. However, a major breakthrough was reported in 2012 when a superconducting one unit cell (UC) thick FeSe thin film was discovered after growing it on top of $SrTiO_3$ (STO) (001) and measuring a large superconducting gap ($\Delta$) of ~20 meV by scanning tunneling spectroscopy (STS).[6] The transition temperature, in that case, was reported to be above 65 K[7] and even up to 109 K[8], which is much higher than in any bulk sample of iron-based superconductors. Interestingly thicker layers (more than one UC) of FeSe on STO were found to be non-superconducting, which is not true for the cases of films grown on bilayer graphene[9] and SiC(0001)[10]. Such unprecedented variations in the $T_C$ of FeSe has generated tremendous interest in the high-$T_C$ community to understand the underlying pairing mechanism for such systems by growing films on various substrates as well as doping them with different carrier concentrations. As an example, potassium intercalated thicker layers ($\geq$ 2UC) of FeSe on STO were found to be superconducting with a large superconducting gap as detected by both ARPES[11] and STS.[12]

On the other hand, several theoretical studies predict that heterostructures of superconducting thin films and topological insulators (TI) can host Majorana fermions[13] at vortex cores due to tunneling of Cooper-pairs into time-reversal invariant surface states via the proximity effect[14]. So far, several studies by STS have been carried out for TI thin films grown on s-wave superconductors like $NbSe_2$ thereby detecting a proximity effect in layers of TI.[15,16] Moreover, a zero bias conductance (ZBC) peak was observed in differential tunneling conductance ($dI/dV$) spectra measured at the vortex core of $Bi_2Te_3$/$NbSe_2$ heterostructures providing evidence for the existence of Majorana fermions.[17] Even in the case of TI/cuprate superconductor, it has been found that superconductivity can be induced in the TI layer[18], but this is not true for the TI on superconducting FeSe/SiC(0001)[10], for which the surface of FeSe becomes non-superconducting.[19] Most of the studies in this respect have focused on the growth of layers of TI on superconductors. There is a recent work, describing a study of the growth of FeSe on $Bi_2Se_3$, but the local electronic structure of the films was not investigated with STS.[20] This motivated us to investigate the electronic and superconducting properties of FeSe grown on TI in order to gain insight into effects of gapless surface states on Cooper-pairs and whether thin FeSe layers can sustain the superconductivity when grown on a TI.

In this work, we first present successful growth studies of one to two UC thick FeSe thin films on top of $Bi_2Se_3$(0001). We show that striped moiré patterns are generated in the films due to the difference between the FeSe and $Bi_2Se_3$ lattice structures and that structural domains of FeSe islands are oriented in three different directions. We discuss STS results obtained on one and two UC thick FeSe islands down to 6.5 K.

# *Experimental details*

Our FeSe/$Bi_2Se_3$ heterostructures were prepared in the following way: $Bi_2Se_3$ single crystals[21] were first cleaved in an ultra-high vacuum environment and then degassed up to 315 °C while maintaining the pressure in the chamber lower than $8 \times 10^{-8}$ mbar. Afterwards, we deposited a few monolayers (ML) of high-purity Fe (99.995%) (the calibration for the deposition was performed on W(110)) from an e-beam evaporator on $Bi_2Se_3$ at room temperature, annealed the sample at ~315 °C for 30 minutes, and subsequently cooled it down to room temperature.

STM measurements were conducted using two home-built ultra-high vacuum STM setups. One of them is a variable-temperature STM with a continuous-flow He cryostat (20K-300K)[22] and the other one is a 4K-STM operating at a base temperature of 6.5 K.[23] Chemically etched tungsten tips were used for both imaging and tunneling spectroscopy. We recorded topographic images in the constant-current mode at the stabilization current $I_t$. The differential tunneling conductance $dI/dV$, which is approximately proportional to the sample's electron density of states at $E_F$, was measured after switching off the feedback using a lock-in amplifier by adding an AC-modulation voltage ($V_{mod}$) at a frequency of 912 Hz to the bias voltage ($V_b$) which is applied to the sample while the tip was at virtual ground.

# *Results and Discussion*

We present an STM topographic image of the bare substrate surface of $Bi_2Se_3$ in Fig. 1(a), which was acquired after annealing the sample up to 320 °C because of the necessity of degassing the glue. The islands which appear on the surface have heights equivalent to multiples of a quintuple layer (QL) of $Bi_2Se_3$ which is ~9.5 Å high.[24,25] Subsequently we deposited 2.5 ML of Fe at room temperature and annealed the sample at 50 °C for 30 minutes. Afterwards, we studied the surface by STM at ≈29 K and obtained the topographic image presented in Fig. 1(b). It can be seen that several types of islands with different heights have grown as represented by the line profile. Further, the same sample was annealed at temperatures of 174 °C [Fig. 1(c)], 242 °C [Fig. 1(d)], and 315 °C [Fig. 1(e)] for the same time period (as mentioned above) and imaged after each annealing step. At higher annealing temperature [242 °C, Fig. 1(d)], elongated features are observable on the surface of $Bi_2Se_3$ indicating that Fe has reacted with the Se-atomic layer of $Bi_2Se_3$. Additionally, islands with adsorbates (bright protrusions) on the top have formed. After annealing at 315 °C [Fig. 1(e)] the surface shows islands with striped patterns which have developed and appeared to be embedded in the $Bi_2Se_3$ since the height difference between the top of the islands and the surface of $Bi_2Se_3$ is on the order of only 1.5 Å - 2 Å (see the line profile presented in the bottom area of the image). A detailed investigation of the atomic structure of this surface with atomically resolved STM images is presented later in the manuscript.

Figure 2(a) shows a topographic STM image of a large area of a sample with a higher nominal Fe coverage with a large number of islands exhibiting a stripe pattern oriented in three different directions. The line profile in Fig. 2(b) shows that these islands exhibit step heights (5.4±0.2) Å and multiples of that. The other surfaces which do not show the stripe pattern are separated by steps with a height of (9.6±0.2) Å. The former value is close to the unit cell height of FeSe and the latter one close to the QL of $Bi_2Se_3$, as can be seen by the schematic diagram in Fig. 2(c).[1,24,25] Upon resolving these different areas atomically, we find that the one with the stripe pattern exhibits a square lattice with a periodicity of $a_{FeSe} \approx (3.7 \pm 0.1)$ Å, indicating that these islands are comprised of the tetragonal phase of FeSe [Fig. 2(e)]. We did not observe the hexagonal phase of FeSe as reported by other STM studies.[26] We can also see dumbbell shaped defects oriented along the *a*- and *b*- crystal axes (or Se-Se bond directions) of the FeSe lattice [marked by dotted green and blue rectangles in Fig. 2(e)]. They may be created due to the replacement of Fe atoms by Se atoms and were seen earlier on the cleaved surface of bulk FeSe[27] as well as thin films.[28] The atomically resolved image of Fig.

2(d) taken on one of the other areas which do not show the stripes reveals a hexagonal lattice-structure as evident from its Fourier transform [inset of Fig. 2(d)]. The measured lattice constant of $a_{\text{Bi}_2\text{Se}_3} \approx (4.1 \pm 0.1)$ Å matches well with the in-plane $a$-axis parameter of $\text{Bi}_2\text{Se}_3$, which is 4.14 Å[25], and is therefore assigned to different QLs of the substrate $\text{Bi}_2\text{Se}_3$ (top and bottom). The defects (dotted circle) appearing with different intensities in the image of Fig. 2(d) have been seen earlier and were attributed to Fe atoms incorporated into a subsurface layer of the $\text{Bi}_2\text{Se}_3$ substrate upon thermal activation [see Ref. 29 for their detailed investigation].

In Fig. 3(a), we have depicted an atomically resolved STM topography showing that both, the $a$-axes as well as the striped patterns of a one UC thin island of FeSe can be oriented in three directions with an angle of 60° in neighboring domains. The stripes can be attributed to a moiré pattern which forms due to the lattice mismatch of the $\text{Bi}_2\text{Se}_3$ and FeSe lattice constants.[20] The three different orientations are traced back to three possible rotational domains of the tetragonal lattice of FeSe on the hexagonal lattice-structure of the $\text{Bi}_2\text{Se}_3$.[20] Figures 3(b) and (c) show the height variation and periodicity of the moiré pattern. The apparent height variation is in the range of 40 pm - 45 pm. The measured periodicity of the moiré pattern ($\delta$) is (7.3±0.1) nm. This suggests a relative orientation of the tetragonal FeSe lattice on the hexagonal $\text{Bi}_2\text{Se}_3$ lattice ($a_{\text{Bi}_2\text{Se}_3} = 4.14$ Å) as shown in Fig. 3(d), i.e. $a_{\text{Bi}_2\text{Se}_3} \cdot \cos 30 \cdot (n+1) = a_{\text{FeSe}} \cdot n = \delta$ with $n \approx 19.4 \pm 0.30$ consistent with Ref. 20. The relation additionally enables an independent estimation of $a_{FeSe}$ resulting in $a_{FeSe} = (3.77 \pm 0.01)$ Å which is close to the bulk lattice constant of $a_{\text{FeSe}} = 3.77$ Å indicating a very little amount of strain in the FeSe thin film. Similar moiré patterns were reported by Wang et al. for the surface of Bi$_2$Se$_3$/FeSe heterostructures[19], and by Cavallin et al.[20] for the same system as investigated here. Note, that, in some of the FeSe areas, we find a considerable misalignment between the FeSe lattice and the Moiré presented in Fig. 3(e), as also reported in Ref. 20. We would like to point out that the growth mode shown in Fig.2, where we observe only one and two UC high FeSe islands *on top* of the Bi$_2$Se$_3$ is different from that shown in the previous study[20] where a few layers of FeSe are *embedded* inside the substrate. Therefore, this sample provides the opportunity to probe the electronic properties of FeSe on Bi$_2$Se$_3$ in the limit of one and two UCs and the corresponding spectroscopic measurements are described in the following.

In order to probe the electronic properties including a potential superconductivity of 1-2 UCs FeSe islands on Bi$_2$Se$_3$(0001), we performed tunneling spectroscopic measurements at low temperature (6.5 K) on a similar sample as the one of Fig.2. The d$I$/d$V$ spectra shown in Fig. 4(b) are measured in an energy window of ±100 mV across $E_F$ and along a line of 4 nm length on a one UC FeSe island (not shown). All individual spectra almost fall on top of each other indicating that the surface is electronically quite homogeneous on this length scale. Note that we could not see any change in the spectra across the boundaries of the moiré pattern. Their asymmetric shape with respect to $E_F$ resembles the spectra that have been taken above $T_C$ on bulk FeSe as shown in Ref. 30. Furthermore, the spectrum shown in Fig. 4(c) is the average of a few spectra measured within a small energy range and over the surface of the STM image of Fig. 4(a) acquired on 1 UC thin island of FeSe. It can be clearly seen that there is a finite density of states at $E_F$ indicating the absence of a superconducting gap. Spectra taken on FeSe islands of two UC thickness are identical to the ones taken on one UC thin islands, as shown in Fig. 4(d) together with a spectrum from the substrate region. Thus our measurements clearly show that both surfaces of one and two UC FeSe islands are electronically homogeneous and not superconducting down to 6.5 K.

Finally, we discuss possible scenarios in order to explain the suppression or lowering of the FeSe superconducting phase transition below our measurement temperature (6.5 K) in the UC thickness limit observed here for the substrate of $Bi_2Se_3$. Superconductivity was recently shown to survive[9] or even be enhanced[6,31] in the thickness limit of a few UCs of FeSe depending on the substrates, i.e. STO,[6] $BaTiO_3$[31] or bilayer graphene[9]. There can be a few possible reasons for the suppression of the superconducting phase: first, the FeSe thin film could be under tensile strain because of the lattice mismatch with the $Bi_2Se_3$ lattice, which would induce a suppression of superconductivity.[32] However, the similar lattice constant for our FeSe film in comparison to bulk FeSe excludes the presence of a significant strain. Second, there might be a considerable charge transfer between the FeSe and the $Bi_2Se_3$ substrate, which would then lead to a shift in the Fermi level of FeSe and thus change the superconducting transition temperature. Our results suggest that this is indeed the case for the FeSe film investigated here. In any case, it would be interesting to investigate the possible formation of antiferromagnetic long-range order in the thin FeSe islands on $Bi_2Se_3$ presented here, which is usually favored in Fe-chalcogenides at the expense of the superconducting phase.[33]

# *Conclusion*

In conclusion, we demonstrated the successful growth of one and two UC thin FeSe films in tetragonal phase on the $Bi_2Se_3(0001)$ topological insulator. We observe striped moiré patterns with a periodicity of ~7.3 nm due to the difference between the FeSe and $Bi_2Se_3$ lattice constants. Our spectroscopic data show that there is no signature of a superconducting gap in d$I$/d$V$ spectra measured over one and two UC thin FeSe islands down to 6.5 K, suggesting that superconductivity is suppressed in ultrathin FeSe films grown on $Bi_2Se_3$.

# *Acknowledgments*

The major part of the project has been funded by the ERC Advanced Grant ASTONISH (No. 338802). JW acknowledge partial support through the DFG priority program SPP1666 (grant No. WI 3097/2).

PS: At the time of submission we learned that a similar work has been reported by A. Eich *et al.*, arXiv:1606.05738 (2016), which differs significantly in terms of growth.

*Corresponding author: usingh@physnet.uni-hamburg.de

§Corresponding author: jwiebe@physnet.uni-hamburg.de

# *Figures*

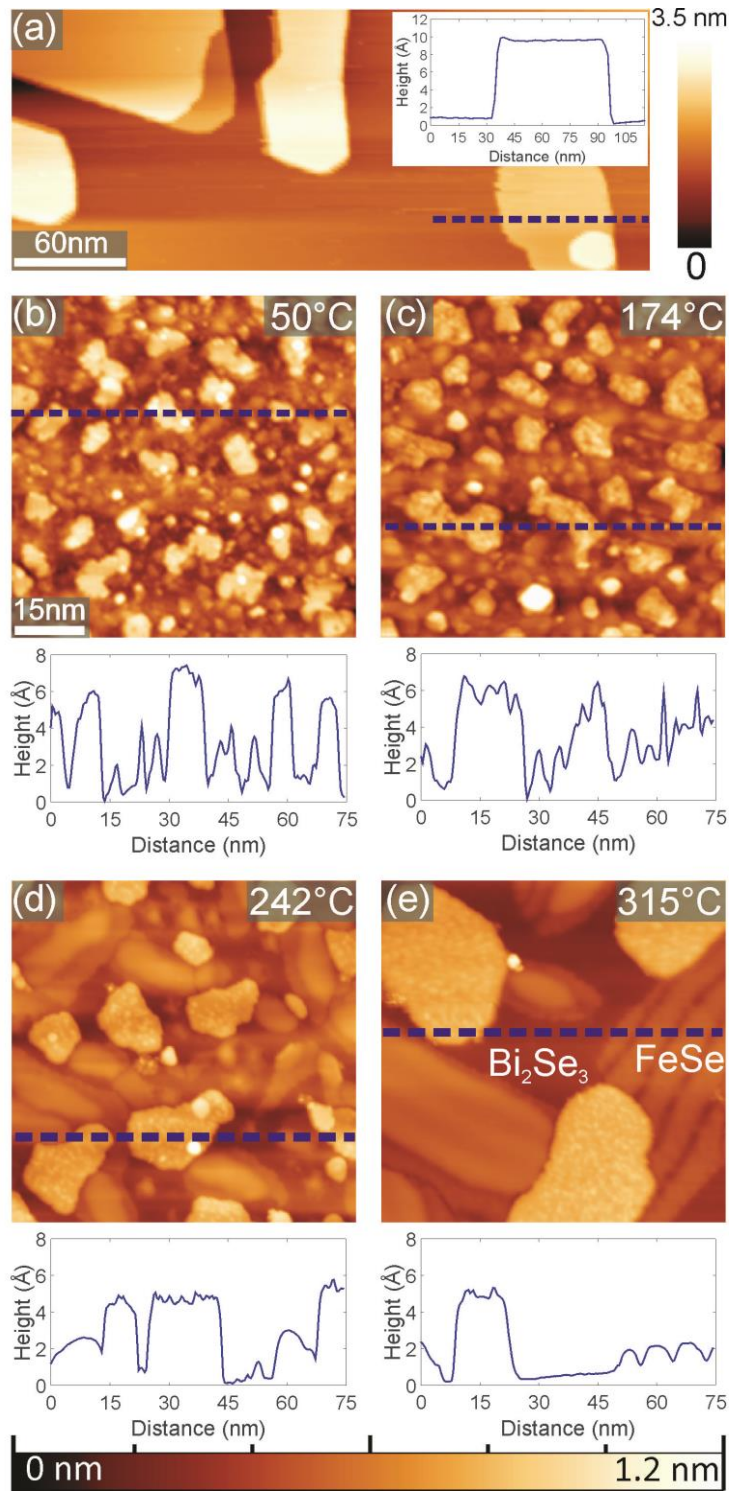

**FIG.1.** (a) Topographic STM image of the bare $Bi_2Se_3$ substrate surface taken after cleavage and degassing the sample up to ~320 °C ($V_b$ = 500 mV, $I_t$ = 10 pA, and $T$ = 30.3 K). The topographies shown in (b), (c), (d), and (e) were measured after depositing 2.5 ML of Fe on $Bi_2Se_3$ and annealing for 30 minutes at temperatures as indicated at the top right of each image (b: $V_b$ = 500 mV, $I_t$ = 10 pA, $T$ = 28.8 K; c: $V_b$ = 500 mV, $I_t$ = 10 pA, $T$ = 28.49 K; d: $V_b$ = 1.0 V, $I_t$ = 10 pA, $T$ = 28.5 K; e: $V_b$ = 1.0 V, $I_t$ = 10 pA, and $T$ = 28.5). Line profiles along the dashed line in each image indicate typical heights of islands formed on the surface.

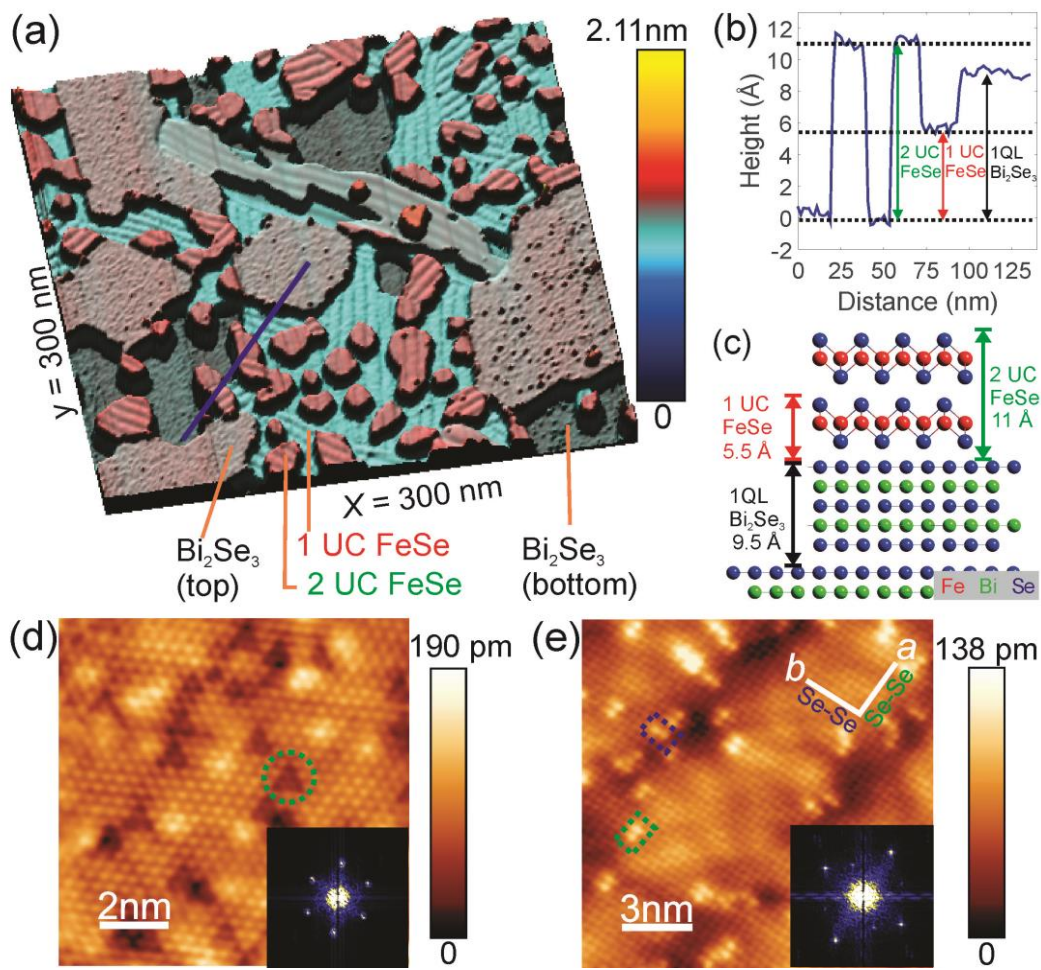

**FIG.2.** (a) 3D view of an STM topographic image of FeSe grown on $Bi_2Se_3$ after deposition of nominally 6 ML Fe and sample annealing at ~305 °C for 30 minutes ($V_b$ = 1 V, $I_t$ = 10 pA, and $T$ = 23.3 K). (b) Line profile along the blue line in (a) showing step heights of 1-2 UCs of FeSe and 1 QL of $Bi_2Se_3$. (c) Schematic diagram showing thicknesses of FeSe UCs and $Bi_2Se_3$ QLs according to literature values [1, 24, 25]. (d) Atomically resolved topographic image of the surface of $Bi_2Se_3$ ($V_b$ = 400 mV, $I_t$ = 100 pA, and $T$ = 26.4 K). The atomic contrast is enhanced by adding raw image data to Fourier filtered data of Bragg peak components in the FFT. The FFT image of (d) is depicted in the inset. The dotted circle denotes a subsurface Fe atom. (e) Atomically

resolved image of the surface of FeSe ($V_b$ = 400 mV, $I_t$ = 100 pA, $T$ = 26.4 K) (FFT image in the inset). Dumb-bell shaped defects marked by rectangles with blue and green colours are oriented along $a$- and $b$- directions corresponding to Se-Se bond directions of the FeSe lattice.

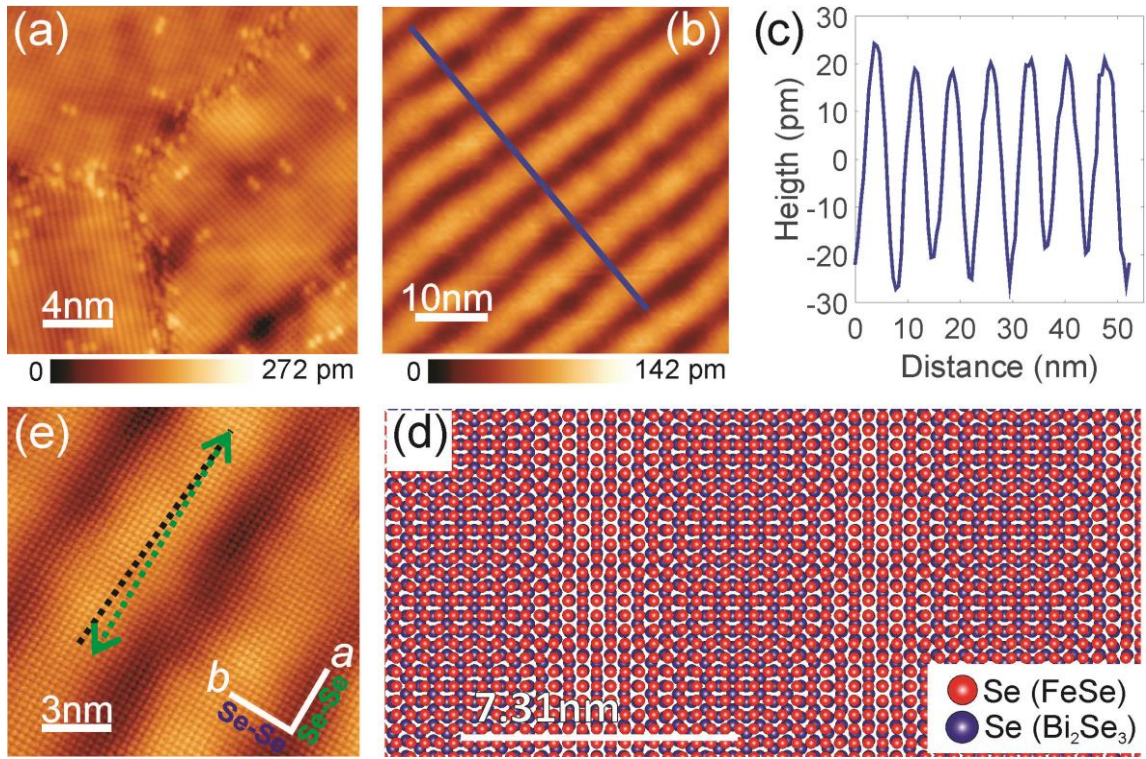

**FIG.3.** (a) Topographic STM image of one UC FeSe on $Bi_2Se_3$ ($V_b$ = 100 mV, $I_t$ = 838 pA, and $T$ = 30.0 K) showing twin boundaries in three different orientations due to the three-fold symmetry of the $Bi_2Se_3(0001)$ lattice. (b) Image ($V_b$ = 4 V, $I_t$ = 10 pA, and $T$ = 27.6 K) of a stripe moiré pattern with a periodicity of (7.3±0.1) nm as evident from a line-profile presented in (c). (d) Schematic diagram of the moiré pattern generated due to the difference between the in-plane lattice constants of FeSe and $Bi_2Se_3$, which are 3.77 Å[1] and 4.14 Å,[24,25] respectively. (e) High-resolution STM data showing a significant misalignment of ~4° between the $a$-axis of the FeSe lattice (green dotted line) and the stripe direction (black dotted line) ($V_b$ = 20 mV, $I_t$ = 500 pA, and $T$ = 28.9 K).

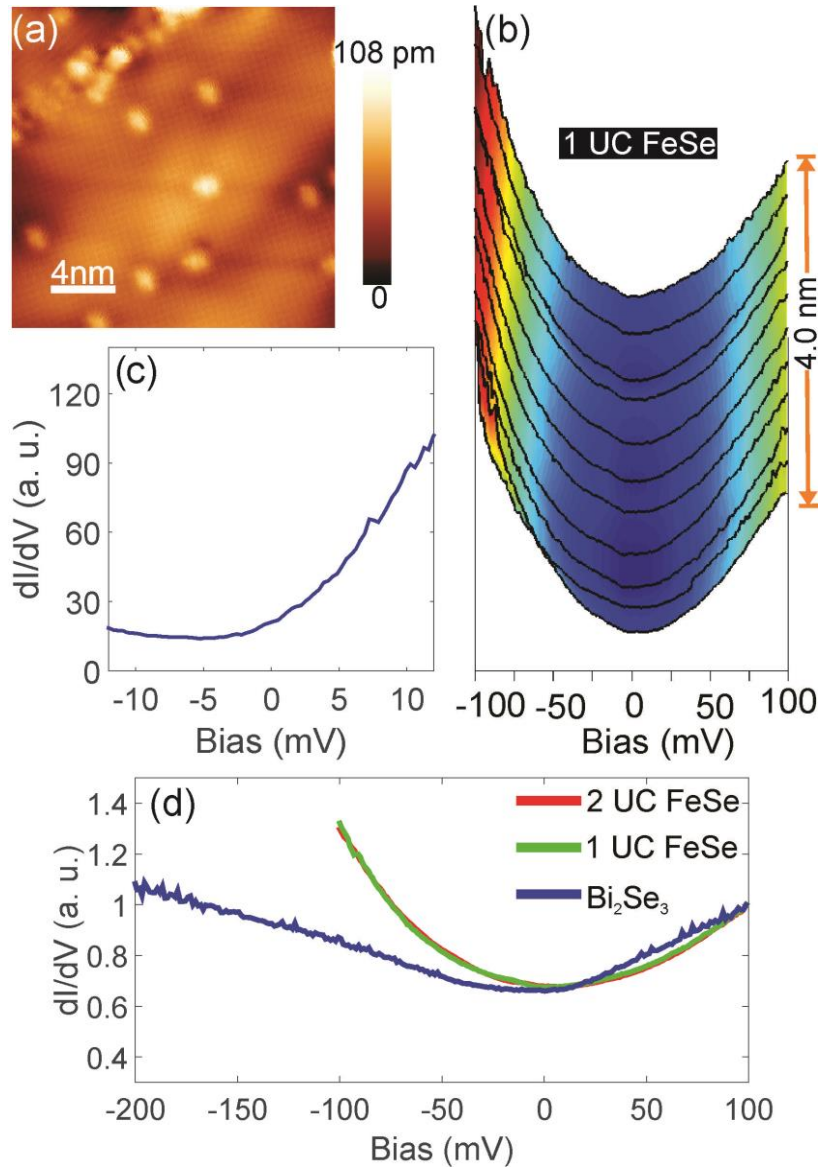

**FIG.4.** (a) Topographic STM image of one UC FeSe on $Bi_2Se_3$ taken at $T = 6.5$ K with $V_b = 200$ mV and $I_t = 300$ pA. (b) Spectra measured along a line profile on one UC thick FeSe island ($V_b = 100$ mV, $I_t = 200$ pA, $V_{mod} = 1.76$ mV, and $T = 6.5$ K). (c) Average of tunneling point spectra measured at $T = 6.5$ K on the surface of image (a) showing no evidence of a superconducting gap ($V_b = 50$ mV, $I_t = 300$ pA, and $V_{mod} = 350$ μV). (d) Spectra measured over a larger energy range on two UC and one UC FeSe with $V_b = 100$ mV, $I_t = 200$ pA, $V_{mod} = 1.76$ mV, and $T = 6.5$ K; and on $Bi_2Se_3$ ($V_b = 100$ mV, $I_t = 250$ pA, and $V_{mod} = 700$ μV, and $T = 6.5$ K). All spectra are normalized to the d$I$/d$V$ value at $V = 100$ mV.